\documentclass[12pt]{article}
\usepackage{graphics}
\usepackage[dvips]{graphicx}
\usepackage{mathrsfs}
\usepackage{amssymb}
\usepackage{verbatim}
\usepackage{float}
\newcommand{\be}{\begin{equation}}
\newcommand{\ee}{\end{equation}}
\newcommand{\bea}{\begin{eqnarray}}
\newcommand{\eea}{\end{eqnarray}}

\def\4vol{{\int d^4x \sqrt{-g}}}

\def\simlt{\stackrel{<}{{}_\sim}}

\newcommand{\nc}{\newcommand}

\nc{\nt}{\tilde{N}}
\nc{\ra}{\rightarrow}
\nc{\lsim}{\begin{array}{c}\,\sim\vspace{-21pt}\\< \end{array}}
\nc{\gsim}{\begin{array}{c}\sim\vspace{-21pt}\\> \end{array}}
\nc{\tnt}{\tilde{N}}
\nc{\tst}{\tilde{t}}

\nc{\LL}{L}
\nc{\vv}{\tilde{v}}

\title{
\vspace*{-1.3cm}
\begin{flushright}
\normalsize{
ANL-HEP-08-76\\
EFI-08-29\\
FERMILAB-PUB-08-561-T\\
MCTP-08-47
}
\end{flushright}
\vspace{0.5cm}
\Large
\textbf{
Minimal Flavor Violation and the Scale of Supersymmetry Breaking
}
\vspace*{0.5cm}
\author{\textbf{M.~Carena$^{a,b}$, A.~Menon$^{c}$ and 
C.E.M.~Wagner$^{b,d,e}$}\\ 
\\[0.5cm]
$^a$\normalsize\emph{Theoretical Physics Dept., Fermi National Laboratory,
Batavia, IL 60510} \\
$^b$\normalsize\emph{EFI and Dept. of Physics, Univ. of Chicago,
5640 S. Ellis Ave., Chicago, IL 60637, USA} \\
$^c$\normalsize\emph{MCTP and Dept. of Physics, University of 
Michigan, Ann Arbor, Michigan 48109, USA} \\
$^d$\normalsize\emph{HEP Division, Argonne National Laboratory,
9700 Cass Ave.,
Argonne, IL 60439, USA} \\
$^e$\normalsize\emph{KICP, Univ. of Chicago,
5640 S. Ellis Ave.,Chicago IL 60637, USA}
}}
\begin{document}
\setcounter{page}{0}
\maketitle
\vspace{0.5cm}

\begin{abstract}
In this paper we explore the constraints from B-physics observables in SUSY 
models of Minimal Flavor Violation, in the large $\tan \beta$ regime, for both 
low and high scale supersymmetry breaking scenarios.  
We find that the rare B-decays  $b \to s \gamma$
and $B_s \to \mu^+ \mu^-$ can be quite sensitive to the scale $M$ at which 
supersymmetry breaking is communicated to the visible sector. In 
the case of high scale supersymmetry breaking, we show that the
additional gluino contribution to the $b \to s\gamma$ and $B_s \to \mu^+ \mu^-$
rare decay rates can be significant for large $\tan \beta$, $\mu$ and $M_3$.
The constraints on $B_u \to \tau \nu$ are relatively insensitive to the
precise scale of $M$. We also consider the 
additional constraints from the present direct Higgs searches at the Tevatron 
in the inclusive $H/A \to \tau \tau$ channel, and the latest CDMS direct 
dark matter detection experiments. We find that altogether the constraints from
B-physics, Higgs
physics and direct dark matter searches can be extremely powerful in probing 
regions of SUSY parameter space for low $M_A$ and large $\tan \beta$, leading
to a preference for models with a lightest CP-even Higgs mass close to
the current experimental limit. 
We find interesting regions of parameter space that satisfy all 
constraints and can be probed by Higgs searches at the Tevatron and the LHC
and by direct dark matter searches in the near future.

\end{abstract}
\thispagestyle{empty}

\pagebreak 

\section{Introduction}

    The next few years promise to be extremely exciting for High Energy Physics
because of new results coming from the Tevatron collider, the expected start 
of the LHC and a number of dark matter detection experiments. It is hoped 
that all this experimental data will shed some light on the mechanism of 
electroweak symmetry breaking and possibly on the origin 
of dark matter in the universe. 

    Theoretically, one of the more promising scenarios that can explain both 
questions is that of low energy supersymmetry. In particular, the minimal 
supersymmetric extension of the Standard Model (MSSM) with R-parity 
can both stabilize the electroweak scale and provide a cold dark matter
candidate (i.e. the lightest neutralino) with a relic abundance that is in 
good agreement with the WMAP value~\cite{Spergel:2006hy}
\bea
\Omega_{CDM} h^2 = 0.105^{+0.007}_{-0.010}. 
\eea
However, like most extensions of the Standard Model, the MSSM is highly 
constrained by flavor changing effects, in particular through 
B-physics observables. 
These constraints can be naturally satisfied if the SUSY breaking 
terms are approximately flavor diagonal at the scale $M$, at which 
supersymmetry breaking is communicated to the visible sector, and all flavor 
changing effects are loop 
induced and proportional to the elements of the CKM matrix of the Standard 
Model. Such supersymmetric extensions of the Standard Model are generically 
called Minimal Flavor Violating (MFV) and have been extensively studied in 
Refs.~\cite{Bertolini:1990if}--\cite{Foster:2005wb}. In particular, in 
Ref.~\cite{Ellis:2007kb} the impact of maximal CP-violation and minimal flavor 
violating MSSM is considered. At large $\tan \beta$, the
ratio of the two Higgs vacuum expectation values in the MSSM, 
flavor changing neutral currents (FCNCs) are induced by the Higgs sector 
through loop effects that can lead to significant deviations 
in B-physics observables from their Standard Model predictions. The 
B-factories, Belle and Babar, and the Tevatron have measured many of these
observables and these data put strong constraints on the allowed MSSM parameter
space.

Simultaneously, experiments are also trying to discover the footprint of 
supersymmetry through dark matter searches of a stable neutralino. These 
searches for dark matter have also begun to put significant constraints on 
supersymmetric models by providing limits on the spin-independent scattering
cross-sections of the lightest neutralino with nuclei. In the
MSSM, the couplings of the down type quarks to the non-standard Higgs bosons 
are $\tan \beta$ enhanced. Therefore the t-channel Higgs boson contribution 
to this cross-section can be sufficiently enhanced for small enough
values of the non-standard CP-even Higgs boson mass and large $\tan \beta$. 
In addition the spin-independent cross-section also depends on the size of the
Higgsino component of the lightest neutralino which is governed by the Higgsino
mass parameter $\mu$. The impact of direct dark matter searches on Higgs 
physics has been analysed in Refs.~\cite{dmdetect, Carena:2006dg, 
Carena:2006nv}.

In this article we study the effect of varying the scale $M$, at which 
supersymmetry breaking is communicated to the visible sector, on B-physics 
observables in the context of Minimal Flavor Violation. We concentrate
on two scenarios
of supersymmetry breaking: low-scale, 
$M \sim M_{SUSY}$, and high-scale, $M \simeq 
M_{GUT}$, SUSY breaking, where $M_{SUSY}$ and $M_{GUT}$
represent the scale of the supersymmetric particle masses 
and the Grand Unification scale, respectively. 
In the case of low scale
supersymmetry breaking the flavor changing effects are governed by loop 
induced Higgs mediated currents. In the case of high scale SUSY breaking, 
the soft squark mass terms are logarithmically sensitive to the  
scale $M$, due to their RG evolution. In particular 
for large supersymmetry breaking scales the soft squark mass parameters 
pick up off-diagonal contributions proportional to the CKM matrix elements. 
Hence the squark and quark mass matrices cannot be diagonalized simultaneously.
This mismatch between the
quark and squark mass bases induces flavor violating quark-squark-gluino 
couplings that 
are proportional to the CKM matrix elements, which lead to important gluino 
contributions to both the $B_s \to \mu^+ \mu^-$ and $b \to s\gamma$ rare
decays, in addition to those already present when $M \sim M_{SUSY}$.

In order to analyze the size of the possible gluino effects, we shall study
scenarios that parametrize the possible flavor violation
effects in models of Minimal Flavor Violation with a small messenger
scale $M$, of the order of the weak scale, and with a large scale $M$,
of the order of the GUT scale, respectively. In the first scenario,
we shall assume no flavor violating quark-squark-gluino couplings. 
In the second scenario, we shall assume a left-handed squark mass
matrix that is diagonalized together with the up Yukawa coupling
matrix, as would be the case if the down Yukawa
effects in the RG evolution of the soft masses were neglected 
compared to those of the up Yukawas.  
The effect of the non-diagonal left-handed down type
quark-squark-gluino vertices on the 
$B_s \to \mu^+ \mu^-$ rare decay within this approximation 
has been previously computed in 
Ref.~\cite{Dedes:2002er}. In this article we derive an 
analytic formula 
for the gluino contribution to the $b \to s\gamma$ rare decay for 
large values of $\tan \beta$, within the same approximation. The
validity of this approximation will be discussed in section~\ref{M=MGUT}. 

In addition we also study the interplay between the B-physics constraints from 
the $B_u \to \tau \nu$, $B_s \to \mu^+ \mu^-$ and the $b \to s\gamma$ rare 
decays and the recent direct dark matter detection limits from 
CDMS~\cite{Ahmed:2008eu}.
Let us stress here that low energy SUSY breaking 
scenarios lead to a light gravitino and therefore the CDMS constraints would 
not apply.
We find that combining the limits from B-physics observables, dark matter 
detection experiments at CDMS and inclusive $H/A \to \tau \tau$ searches at 
the Tevatron~\cite{tevHA} yields interesting constraints on the $M_A \-- \tan 
\beta$ and $X_t\-- \mu$ plane, where $M_A$ is the CP-odd Higgs mass and $X_t$ 
is the stop left-right
mixing parameter. We find regions of parameter space that 
satisfy all these constraints and
can be probed by Higgs searches at the Tevatron and by 
direct dark matter searches in the near future.

The paper is organized as follows, in Section~\ref{theory:sec} we consider the 
effect of the scale $M$ on B-physics observables in minimal flavor violating 
MSSM. 
In particular we present the additional gluino contributions to the $B_s \to 
\mu^+ \mu^-$ and $b \to s\gamma$ rare decays. The complete calculation 
of the gluino contribution to the $b\to s\gamma$ decay can be found in 
Appendix~\ref{append:1}. We also give a brief
theoretical overview of the relevant direct dark matter detection 
cross-section.
In Section~\ref{numerical:sec} we consider different parametric scenarios that 
can satisfy the B-physics experimental constraints, the limits coming from 
inclusive $H/A \to \tau \tau$ searches at the Tevatron and the direct dark 
matter detection limits from CDMS. In particular we explore the dependence of 
our results on the scale at which supersymmetry breaking is communicated to
the visible sector. In Section~\ref{concl:sec} we present our conclusions.

\section{Basic Theoretical Setup} \label{theory:sec}

\subsection{B-physics Constraints and messenger mass scale $M$}

The FCNCs induced by loops of squarks depend on the flavor structure of the 
soft squark mass parameters which, in MFV, is closely tied to the scale at 
which 
supersymmetry breaking is communicated to the visible sector~\footnote{ Unlike
Ref.~\cite{Cohen:2006qc}, we
are considering the case where effects from the hidden sector are small}.  
Assuming the squark masses are flavor independent at high 
scales, the only one-loop corrections that violate flavor are due to RG effects
governed by the up and down Yukawa matrices, since the gauge interactions are 
flavor blind. The corrections to the left-handed soft SUSY breaking mass parameter 
at one-loop are given by~\cite{Dugan:1984qf}
\begin{eqnarray}
\Delta M_{\tilde{Q}}^2 &\simeq& - \frac{1}{8\pi^2}\left[
\left(M_{\tilde{Q}}^2 +  M_{\tilde{u}_R}^2
+ M_{H_u}^2(0) + A_0^2 \right) Y_u^{\dagger} Y_u + \right. \nonumber \\
& & \left.
\left(M_{\tilde{Q}}^2 + M_{\tilde{d}_R}^2
+ M_{H_d}^2(0) + A_0^2 \right) Y_d^{\dagger} Y_d \right] 
\log\left(\frac{M}{M_{SUSY}}\right), \label{mQl2:eq}
\end{eqnarray}
where $M_{\tilde{Q}}^2$ denotes the left-handed squark mass matrix, 
$M_{\tilde{u}_R}^2 (M_{\tilde{d}_R}^2)$ is the right-handed up (down) squark 
mass matrix, $M_{H_{u,d}}^2(0)$ and $A_0$
are the Higgs soft supersymmetry breaking and squark-Higgs 
trilinear mass parameters, respectively, at the messenger scale $M$, 
at which supersymmetry breaking is transmitted to the observable sector, and 
$M_{SUSY}$ is 
the characteristic low energy squark mass scale. Similarly,
the right-handed up and down squark mass matrices, receive
one-loop Yukawa-induced corrections proportional to
\begin{equation}
\Delta M_{\tilde{u}_R}^2 = - \frac{2}{8\pi^2} \left(M_{\tilde{Q}}^2+  
M_{\tilde{u}_R}^2 + M_{H_u}^2(0) + A_0^2 \right) Y_u Y_u^{\dagger}
\log\left(\frac{M}{M_{\rm SUSY}}\right) ,
\end{equation}
and
\begin{equation}
\Delta M_{\tilde{d}_R}^2 = - \frac{2}{8\pi^2} \left( M_{\tilde{Q}}^2+ 
M_{\tilde{d}_R}^2 + M_{H_d}^2(0) + A_0^2 \right) Y_d Y_d^{\dagger}
\log\left(\frac{M}{M_{\rm SUSY}}\right) ,
\end{equation}
respectively. Hence the corrections to the right-handed soft mass parameters
are diagonal in the quark basis, but the left-handed soft mass parameters
of the down squarks pick up off-diagonal contributions proportional to the CKM 
matrix elements.
The size of these corrections depends on the scale $M$ at which SUSY breaking 
is communicated to the visible sector. If $M$ is of the order of $M_{SUSY}$ 
then these corrections are small and if $M \simeq M_{GUT}$ then these 
corrections can be substantial. In this section we consider the effect of these
two scenarios on three B-physics processes $b \to s\gamma$, $B_u \to \tau \nu$ 
and $B_s \to \mu^+ \mu^-$.

\subsubsection{$M \sim M_{SUSY}$ }
In the case $M \sim M_{SUSY}$ the squark mass matrices
are approximately block diagonal which leads to all the neutral Higgs induced 
FCNCs being proportional to the chargino-stop loop 
factor $h_t^2 \epsilon_Y$, with~\cite{Buras:2002vd}
\bea
\epsilon_Y \approx \frac{1}{16 \pi^2} A_t \mu C_0(m_{\tilde{t}_1}^2,
m_{\tilde{t}_2}^2,|\mu|^2) \label{eY:eq}
\eea
where
\bea
C_0(x,y,z) = \frac{y}{(x-y)(z-y)} \log(y/x) + \frac{z}{(x-z)(y-z)} \log(z/x) .
\eea
No flavor changing effects are produced by contributions from the 
gluino down squark loop as they are purely flavor diagonal
\bea
\epsilon_0^{I} &\approx& \frac{2 \alpha_s}{3 \pi} M_3 \mu C_0(
m_{\tilde{d}_{I,1}}^2,m_{\tilde{d}_{I,2}}^2,|M_3|^2), \label{e0j:eq}
\eea
where $m_{\tilde{d}_{I,1}}$ and $m_{\tilde{d}_{I,2}}$ are the I$^{th}$ down
squark mass eigenstates. The effective flavor changing 
strange-bottom-neutral-Higgs coupling 
is~\cite{Buras:2002vd,Demir:2003bv,Foster:2005wb}
\bea
(X_{RL}^S)^{32}=\frac{\bar{m}_{b} y_t^2 \epsilon_Y 
(x_u^S - x_d^S \tan \beta)}{v_d (1+
\epsilon_0^3 \tan \beta) (1+\epsilon_3 \tan \beta)} V_{eff}^{33*} 
V_{eff}^{32} \label{xrl:eq}
\eea
where 
\bea
\epsilon_3 &=& \epsilon_0^3 +y_t^2 \epsilon_Y \label{e3:eq}\\
x_d^S &=& (\cos \alpha,-\sin \alpha,i\sin \beta) \\
x_u^S &=& (\sin \alpha, \cos \alpha, -i\cos \beta)
\eea 
in the basis ($S={H^0,h^0,A^0}$). 

At large values of $\tan \beta$ the dominant 
supersymmetric contributions to rare decay $B_s \to \mu^+ \mu^-$ are mediated 
by neutral Higgs boson exchange that leads to~\cite{Buras:2002vd,Carena:2006ai}
\bea
\mathcal{BR}(B_s\rightarrow \mu^+ \mu^-) = 4.64 \times 10^{-6} M_{B_s}^2 \left(
\frac{4 \pi^2 m_{\mu} \tan \beta}{\bar{m}_b M_W^2 2^{7/4} G^{3/2} |V_{eff}^{ts}
|}\right)^2 \frac{|(X_{RL}^A)^{32}|^2}{M_A^4}.\label{bstbma:eq}
\eea
Therefore, in this scenario, the magnitude of this observable is suppressed 
when $|\mu A_t|$ is small compared to $M_{SUSY}^2$.

As the gluino-quark-squark vertex is flavor diagonal for $M \sim M_{SUSY}$ the 
dominant SUSY contributions to the $b \to s \gamma$ rare decay come from the 
charged-Higgs boson and the chargino-stop loops. In particular the Wilson 
coefficients due to the charged Higgs contribution are proportional to the 
factor~\cite{Degrassi:2000qf,Carena:2000uj}
\begin{eqnarray}
C^{H+}_{7,8} \propto \frac{h_t - \delta h_t}{1+\epsilon_3 \tan \beta}, 
\label{bsgh+:eq}
\end{eqnarray}
while the Wilson coefficient due to the chargino-stop loop has the 
form~\cite{Degrassi:2000qf,Carena:2000uj}
\begin{eqnarray}
C^{\chi}_{7,8} \propto \frac{\mu A_t \tan \beta}{1+ \epsilon_3 \tan \beta} 
f(m_{\tilde{t}_1}^2,m_{\tilde{t}_2}^2,m_{\chi^+}^2) \label{bsgchi:eq}
\end{eqnarray}
where $f$ is the loop integral appearing at one loop. Eq.~(\ref{bsgh+:eq})
includes the $\tan \beta$ resummed contributions and the prescription used in 
Refs.~\cite{Degrassi:2000qf,Carena:2000uj},
\bea
& & h_t \rightarrow h_t - \delta h_t \\
& & m_b \rightarrow \frac{m_b}{1+\epsilon_3 \tan \beta} \\
& & \delta h_t = \frac{2 \alpha_s}{3\pi} \mu M_3 \tan \beta
\left(\cos^2 \theta_{\tilde{t}} C_0(m_{\tilde{s}_L}^2,m_{\tilde{t}_1}^2,M_3^2) 
+ \sin^2 \theta_{\tilde{t}} C_0(m_{\tilde{s}_L}^2,m_{\tilde{t}_2}^2,M_3^2) 
\right),
\eea
where, $\delta h_t$ is the correction to the charged-Higgs-top-strange vertex
due to the gluino-stop loop and $\theta_{\tilde{t}}$ is the stop mixing angle.
 
The dominant supersymmetric contribution to the $B_u \to \tau \nu$ rare decay 
is due to the charged Higgs which interferes with the Standard Model 
contribution and we can define the ratio~\cite{btnu}
\bea
R_{B\tau \nu} = \frac{\mathcal{BR}(B_u \to \tau \nu)^{\rm MSSM}}{\mathcal{BR}(
B_u \to \tau \nu)^{\rm SM}} = \left[1 - \left(\frac{m_B^2}{m_{H^{\pm}}^2} 
\right) \frac{\tan^2 \beta}{1+\epsilon_0 \tan \beta}\right]^2. \label{rbtnu:eq}
\eea
so as to quantify deviations from the Standard Model in this process.

In addition, Ref.~\cite{Antonelli:2008jg} has shown the importance
of Kaon semi-leptonic decays in constraining the charged Higgs contribution 
to the $B_u \to \tau \nu$ rare decay. In particular they consider the quantity 
\bea
R_{l23} = \left| \frac{V_{us}(K_{l2})}{V_{us}(K_{l3})} 
\frac{V_{ud}(0^+ \to 0^+)}{V_{ud}(\pi_{l2})}\right|
\eea
where the subscript $li$ refers to semileptonic decays with $i$ final states 
and $0^+ \to 0^+$ refers to beta decay. For the Standard Model,
$R_{l23} = 1$ while when a charged Higgs is included we have
\bea
R_{l23} = \left|1-\frac{m_K^2}{m_{H^{\pm}}^2} \left(1-\frac{m_d}{m_s}\right) 
\frac{\tan^2 \beta}{1+\epsilon_0 \tan \beta} \right| \label{rl23:eq}.
\eea
The charged Higgs constribution in Eq.~(\ref{rl23:eq}) and
Eq.~(\ref{rbtnu:eq})
are the same and limits on $R_{l23}$ can be a strong constraint on the
scenario in which SUSY contributions to the $B_u \to \tau \nu$ dominate those 
of the Standard Model. Assuming that $\delta$ is the largest allowed negative 
deviation 
of $R_{l23}$ from one and $\xi$ is the smallest allowed value of 
$R_{B\tau \nu}$, we see that for the charged Higgs to dominate over 
the SM contributions in Eq.~(\ref{rbtnu:eq}) the deviations must satisfy
the constraint
\bea
\delta \geq \frac{m_{K^+}^2}{m_{B_u}^2} \left(1-\frac{m_d}{m_s}
\right) (1+\sqrt{\xi}) \approx 0.008 (1+\sqrt{\xi}) \label{rl23_delta:eq}
\eea
Hence, a two sigma experimental bound on  
$\delta \lsim 0.008 (1 + \sqrt{\xi})$ would strongly disfavor scenarios in 
which the charged Higgs contribution to the $B_u \to \tau \nu$ decay is larger
than that of the Standard Model.

\subsubsection{$M \simeq M_{GUT}$}
\label{M=MGUT}

When $M \simeq M_{GUT}$, corrections to the soft masses due to RG evolution
are log enhanced. Therefore, if we neglect the $Y_d^{\dagger} Y_d$ term
in Eq.~(\ref{mQl2:eq}), the left-handed down squark
mass matrix is diagonalized by the matrix $U_L$ which 
diagonalizes the up-quark mass matrix, rather than the down quark 
diagonalizing 
matrix $D_L$. Neglecting the corrections due to the bottom Yukawa 
over-estimates the splitting between the third and first two generations of 
down squark masses and is not valid when $y_b \sim y_t$ or when $\tan \beta$ is
large.  For $\mu$ and $M_3$ of the order of $M_{SUSY}$, with $\mu M_3$ 
positive, one obtains $\epsilon_3 \sim 0.01$ and therefore the bottom Yukawa, 
\bea
y_b = \frac{m_b \tan \beta}{v(1+\epsilon_3 \tan \beta)}
\eea 
becomes equal to $y_t$ for values of $\tan \beta \gsim 100$. 
The parametrization used in this article increases in accuracy as
$\tan\beta$ takes smaller values, and also for larger values of 
$\mu$, for which the above corrections to the bottom Yukawa coupling
become significant, therefore reducing the value of $y_b$. 
In this article, we shall assume 
that $\tan \beta \lsim 60$. In addition, as we shall discuss below, 
present experimental constraints lead to a preference for moderate or large
values of $\mu$ at sizable values of $\tan\beta$ and small values
of $M_A$. Therefore, we expect our parametrization to lead to a good 
approximation of the gluino induced effects in the scenarios discussed in this 
article. Furthermore using this approximation we were able to reproduce the the
numerical B-physics limits obtained by Ref.~\cite{Ellis:2007ss}, where the full
renormalization group evolution of the mass parameters was performed.

In the approximation, in which the left-handed down squarks are 
diagonalized by 
$U_L$, flavor violating vertices proportional to the 
CKM matrix in the gluino-down squark-down quark interaction
vertex are induced,
\bea
\mathcal{L}_g &\supset& \sqrt{2} g_3 \tilde{g}^a \left( (V_{CKM})^{JI} 
(\tilde{d}_L^{*})^J T^a d_L^I - (\tilde{d}_R^{*})^I T^a d_R^I \right), 
\label{gluino_nmfv:eq}
\eea
and the soft SUSY breaking down-squark mass Lagrangian takes the form
\bea
\mathcal{L}_{mass} &\supset& (\tilde{d}_L^*)^I (m_Q^2)^{I} (\tilde{d}_L)^J + 
(\tilde{d}_R^*)^I 
(m_R^2)^{I} (\tilde{d}_R)^J + \tilde{\mu}^* (\tilde{d}_L^*)^I 
V_{CKM}^{IJ} m_{d_J} (\tilde{d}_R)^I + h.c. \label{mass_nmfv:eq}
\eea
where $\tilde{\mu} = \mu \tan \beta - A_b$.
Due to the gluino-quark-squark couplings being non-diagonal there are 
additional contributions to both the loop induced $B_s \to \mu^+ \mu^-$ and 
$b \to s\gamma$ rare decays, but no large additional contributions to the 
$B_u \to \tau \nu$ process. 

For $M \simeq M_{GUT}$, the effective flavor changing 
strange-bottom-neutral-Higgs coupling is~\cite{Dedes:2002er}
\bea
(X_{RL}^S)^{JI}=\frac{\bar{m}_{d_J} (\epsilon_3-\epsilon_0) (x_u^S - x_d^S 
\tan \beta)}{v_d (1+\epsilon_0 \tan \beta) (1+\epsilon_3 \tan \beta)} 
V_{eff}^{3J*} V_{eff}^{3I} \label{xrlnmfv:eq}
\eea
where we have assumed that the first two generations of left-handed down 
squark masses are $m_{0}$, the uniform right-handed down 
squark soft mass parameters are $m_R$ and
\bea
\epsilon_0 &\approx& \frac{2 \alpha_s}{3 \pi} M_3 \mu C_0(
m_{0}^2, m_{R}^2,|M_3|^2).
\eea
In the limit of the left-handed sbottom mass being equal to that of 
the first two generations, Eq.~(\ref{xrlnmfv:eq}) reduces 
to Eq.~(\ref{xrl:eq}). In the
$M \simeq M_{GUT}$ scenario, the dominant SUSY contribution to 
$B_s \to \mu^+ \mu^-$ rare decay, at large $\tan \beta$, is found by 
substituting the form of $X_{RL}^{32}$ in Eq.~(\ref{xrlnmfv:eq}) into 
Eq.~(\ref{bstbma:eq}). The present experimental limit on $\mathcal{BR}(B_s \to 
\mu^+ \mu^-)$ disfavors very large positive contributions due to
new physics effects. In high scale SUSY breaking, 
the new physics contributions to  the $B_s \to \mu^+ \mu^-$ 
rare decay process are suppressed if the splitting in the 
left-handed down-type squarks soft mass parameters is such 
that $\epsilon_3-
\epsilon_0$ is rendered small. As $|\epsilon_0| <|\epsilon_0^3|$ this 
suppression may be significant whenever  $\mu A_t < 0$, where 
the value of $|\mu A_t|$, that allows such a cancellation, increases with the 
splitting of down squark masses and therefore with the messenger mass scale
$M$.

Futhermore, flavor violation in the gluino sector also induces relevant
gluino contributions to the $b \to s \gamma$ rare 
decay~\cite{Dudley:2008dp,Wick:2008sz}. In 
Appendix~\ref{append:1} we find that, within the approximation of 
Eq.~(\ref{gluino_nmfv:eq}) and Eq.~(\ref{mass_nmfv:eq}), the Wilson 
coefficients due to these gluino contributions are
\bea
C_{7,8}^{\tilde{g}} &=& \frac{\sqrt{2} \pi \alpha_s}{G_f} (m_{0}^2-
m_{Q_3}^2) \frac{M_3 e^{-i\phi}}{m_b} \left(\frac{f_{\gamma,g}^5(
x_{g0})}{m_{0}^2} \frac{|\tilde{\mu}|m_b}{(m_{0}^2-m_{b_1}^2)(
m_{0}^2-m_{b_2}^2)} \right.\\
& & \left. + s_{\theta} c_{\theta} \left\{\frac{f_{\gamma,g}^5(x_{g1})}{
m_{b_1}^2 (m_{b_1}^2-m_{0}^2)}- \frac{f_{\gamma,g}^5(x_{g2})}{m_{
b_2}^2(m_{b_2}^2-m_{0}^2)}\right\} \right) \nonumber
\eea
where $\cot {2\theta} = \left(m_{Q_3}^2 - m_R^2 \right)/(2|\tilde{\mu}| m_b)$,
$m_{Q_3}$ is the 
left-handed third generation down squark mass, $m_{b_i}$ is the
$i^{th}$ sbottom mass, $\tilde{\mu} = \mu \tan \beta - A_b$, $\phi = 
\arg(\tilde{\mu})$, $x_{gi} = M_3^2/m_{b_i}^2$ and the $f^{i}_{\gamma,g}$ functions
are defined in Eq.~(\ref{fs:eq})~\footnote{
A calculation of the gluino effects valid in the more general case
has been recently performed
in Ref.~\cite{Degrassi:2006eh}}. As expected this contribution to $b \to s
\gamma$ rare decay also disappears in the limit of uniform left-handed down
squark soft mass parameters $m_{Q_3}=m_{0}$. For non-zero mass splittings, 
these contributions are important at large $\tan \beta$ and in the absence of 
CP violation are proportional to the 
sign of $\mu M_3$. Therefore if $\mu M_3$ is
positive the gluino contribution adds to that of the charged Higgs while
when it is negative it subtracts from the charged Higgs contribution.  

\begin{figure}
\begin{center}
\resizebox{9.cm}{!}{\includegraphics{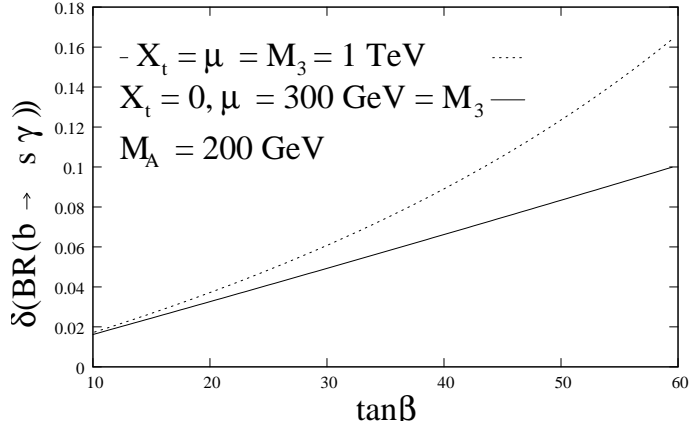}}
\end{center}
\caption{Variation of the gluino constribution to the $b \to s\gamma$ rare
decay branching ratio 
as a function of $\tan \beta$ for two different sets of SUSY parameters, 
assuming a  20\% splitting in the squark masses and a value of $M_A=200$~GeV.}
\label{dbsgvstb:fig}
\end{figure}

In Fig.~\ref{dbsgvstb:fig} we plot the relative gluino contribution to the
decay branching ratio
$\mathcal{BR}(b \to s\gamma)$, 
defined as
\bea
\delta (\mathcal{BR}(b \to s\gamma)) = \frac{\mathcal{BR}(b \to s\gamma)_{\rm with \; 
gluinos} - \mathcal{BR}(b \to s\gamma)_{\rm without \; gluinos}}{\mathcal{BR}(b
\to s\gamma)_{\rm without \; gluinos}},
\eea
for two different sets of SUSY parameters.
The solid curve corresponds to the value of $\delta(\mathcal{BR}
(b \to s\gamma))$ for 
superymmetric parameters $\mu = M_3 = 300$~GeV and $X_t = 0$, while for the 
dashed curve we consider
$\mu = M_3 = -X_t =1$~TeV. The splitting between the third
and first two generations of squark masses is 20\%  and $M_A \sim 200$~GeV.
We see that, in general, the gluinos lead to a moderate modification
of $\mathcal{BR}(b \to s \gamma)$. For instance, in the example shown 
in Fig.~\ref{dbsgvstb:fig},  
the gluino effects lead to  at most a 10-15\% contribution to 
$\mathcal{BR}(b \to s \gamma)$.
In addition, since $\mu M_3$ is positive for these points, 
the gluino contribution to the $\mathcal{BR}(b \to s\gamma)$ is also positive.

\subsection{Direct dark matter detection through Higgs exchange} \label{dmth:sec}

The spin-independent elastic scattering cross-section for a neutralino 
scattering off a heavy nucleus is:
\bea
\sigma_{SI} = \frac{4m_r^2}{\pi} \left(Z f_p + (A-Z) f_n\right)^2
\eea
where $m_r = \frac{m_N m_{\chi^o}}{m_N+m_{\chi^o}}$, $m_N$ is the mass of the
nucleus, $m_{\chi^o}$ is the neutralino mass,
\bea
f_{p,n} &=& \left(\sum_{q=u,d,s} f_{T_q}^{(p,n)} \frac{a_q}{m_q} + \frac{2}{27}
f_{TG}^{(p,n)} \sum_{q=c,b,t} \frac{a_q}{m_q} \right) m_{p,n} \\
a_u &=& - \frac{g_2 m_u}{4m_W s_{\beta}} (g_2 N_{12} - g_1 N_{11}) 
\left[N_{13} s_{\alpha} c_{\alpha} \left( \frac{1}{m_h^2} - \frac{1}{m_H^2}
\right) + N_{14} \left(\frac{c^2_{\alpha}}{m_h^2} +\frac{s^2_{\alpha}}{m_H^2}
\right)\right] \label{au:eq} \\
a_d &=& - \frac{g_2 \bar{m}_d}{4m_W c_{\beta}} (g_2 N_{12} -
g_1 N_{11}) 
\left[N_{14} s_{\alpha} c_{\alpha} \left( \frac{1}{m_h^2} - \frac{1}{m_H^2}
\right) - N_{13} \left(\frac{s^2_{\alpha}}{m_h^2} +\frac{c^2_{\alpha}}{m_H^2}
\right)\right] \label{ad:eq}, 
\eea 
and the quark form factors are $f_{T_u}^p = 0.020 \pm 0.004, f_{T_d}^p = 0.026 
\pm 0.005,  f_{T_s}^p = 0.118 \pm 0.062, f_{TG}^p \approx 0.84, 
 f_{T_u}^n = 0.014 \pm 0.003, f_{T_d}^n = 0.036 \pm 0.008, 
 f_{T_s}^n = 0.118 \pm 0.062$ and $ f_{TG}^n \approx 0.83 $~\cite{dmdetect}. 
In Eq.~(\ref{au:eq}) and Eq.~(\ref{ad:eq}) $N_{ij}$ is the neutralino rotation 
matrix, $\alpha$ is the CP-even Higgs rotation angle and $m_h$ ($m_H$) is the 
lighter (heavier) CP-even Higgs mass. Also in the above expression we define
\bea
\bar{m}_d = \frac{m_d}{1 + \epsilon_0 \tan \beta} 
\eea
for the first two generations of quarks and 
\bea
\bar{m}_b = \frac{m_b}{1 + \epsilon_3 \tan \beta}
\eea
for the bottom quark. In Eq.~(\ref{ad:eq}), we
are ignoring the contribution from s-channel squark exchange, which becomes 
subdominant for heavy squark masses. 
In the limit of large $\tan \beta$ $a_d$ is $\tan \beta$ enhanced 
compared to $a_u$. Moreover, for large $\tan\beta$, $\mu \gg M_1$,
$M_2 \simeq M_1$ and 
$120\mbox{ GeV} \lsim
M_A \lsim  600$~GeV $\;$ 
($M_A \lsim 120$~GeV), one obtains
$N_{11} \gg N_{12}$, 
$m_H \simeq M_A$ $\;$ ($m_h \simeq M_A$) and
$s_{\alpha} \sim -1/\tan \beta$ $\;$
($c_{\alpha} \sim 1/\tan \beta$). Hence we find
that the dominant contribution is
\bea
f_{p,n} &\approx& -m_{p,n} \left(\frac{f_{T_d}^{p,n} + f_{T_s}^{p,n}}{1 + 
\epsilon_0 \tan \beta}  + \frac{2}{27} \frac{f_{TG}^{p,n}}{1 + 
\epsilon_3 \tan \beta}\right) \frac{g_1 g_2 N_{11} N_{13} \tan \beta}{4 m_W 
M_A^2} \\
&\approx& - \left(\frac{0.14}{1 + \epsilon_0 \tan \beta}  + \frac{0.06}{1 + \epsilon_3 \tan \beta}\right) m_p \frac{g_1 g_2 N_{11} N_{13} \tan \beta}{4 
m_W M_A^2}
\eea
where we have neglected, in the first line, the splitting between the first two
generations of squarks and, in the second line, the differences between the 
proton and the neutron mass and we used the fact that the neutron and proton 
$f_T$ factors are relatively similar. Assuming that
the mass of the neutralino is much larger than that of the nucleus we have 
$m_r \sim m_N \sim A m_p$ and
\bea
\sigma_{SI} &\approx& \frac{4A^2 m_p^2}{\pi} A^2 f_p^2 \\
\Rightarrow \frac{\sigma_{SI}}{A^4} &\approx& \frac{g_1^2 g_2^2 N_{11}^2 
N_{13}^2 m_p^4 \tan^2 \beta}{4\pi m_W^2 M_A^4} \left(\frac{0.14}{1 + \epsilon_0 \tan \beta}  + \frac{0.06}{1 + \epsilon_3 \tan \beta}\right)^2, 
\label{sigma_si:eq}
\eea
where $\sigma_{SI}/A^4$ is the neutralino nucleon spin-independent 
cross-section. From Eq.~(\ref{sigma_si:eq}) the spin-independent cross-section 
scales as $\tan^2 \beta/M_A^4$ and therefore direct dark matter detection 
experiments like CDMS can put strong constraints on regions of small $M_A$ and 
large $\tan \beta$.

\begin{figure}
\begin{center}
\resizebox{3.cm}{!}{\includegraphics{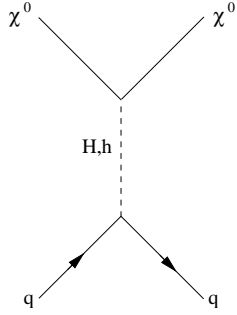}}
\end{center}
\caption{Feynman diagram of the t-channel CP-even Higgs contribution to the 
spin-independent cross-section.}
\label{dm_higgs:fig}
\end{figure}

\section{Numerical limits and constraints} \label{numerical:sec}

\subsection{Experimental constraints on B-physics observables.}

Due to the extra supersymmetric contributions to these rare decays we find
that experimental data put strong constraints on these models.
In particular, the world experimental average of the branching ratio of 
the $b \to s \gamma$ rare decay is~\cite{Barberio:2006bi}
\begin{eqnarray}
\mathcal{BR}(b\to s \gamma)^{exp} = (3.55 \pm 0.24^{+0.09}_{-0.10} \pm 0.03) 
\times 10^{-4},
\label{bsgbound:eq}
\end{eqnarray} 
which agrees well with the Standard Model prediction~\cite{Misiak:2006zs}
\begin{eqnarray}
\mathcal{BR}(b \to s \gamma)^{SM} = (3.15 \pm 0.23) \times 10^{-4}. 
\label{bsgsm:eq}
\end{eqnarray}
Using the experimental and SM ranges for the $b \to s \gamma$ rare decay we 
find the $2\sigma$ allowed range is
\begin{eqnarray}
0.89 \leq  R_{b \to  s \gamma} = \frac{\mathcal{BR}(b \to s \gamma)^{MSSM}}{
\mathcal{BR}(b \to s \gamma
)^{SM}} \leq 1.36. \label{bsgRbound:eq}
\end{eqnarray}

For the $B_u \to \tau \nu$ rare decay the Belle experimental collaboration 
measures a branching ratio of~\cite{Ikado:2006un}
\begin{eqnarray}
\mathcal{BR}(B_u \to \tau \nu)^{\rm Belle} = (1.79^{+0.56}_{-0.49}(\mbox{stat})^{
+0.46}_{-0.51}({\rm syst}))
\times 10^{-4} \label{butaunuexp:eq},
\end{eqnarray}
while the Babar collaboration finds the preliminary value~\cite{Aubert:2006fk}
\begin{eqnarray}
\mathcal{BR}(B_u \to \tau \nu)^{\rm Babar} = (1.20 \pm 0.54) \times 10^{-4}.
\end{eqnarray}
The average of these two experiments is then~\cite{Bona:2006ah}
\begin{eqnarray}
\mathcal{BR}(B_u \to \tau \nu)^{\rm Exp} = (1.41 \pm 0.43 ) \times 10^{-4}.
\end{eqnarray}
Using $f_B= 189 \pm 27$~MeV from LQCD~\cite{Bona:2006ah} and the 
average value of $|V_{ub}| = (3.98 \pm 0.45) \times 10^{-4}$ from 
HFAG~\cite{HFAG}, the Standard Model prediction is 
\bea
\mathcal{BR}(B_u \to \tau \nu) = (1.09 \pm 0.40) \times 10^{-4}
\eea
Assuming at most a 2$\sigma$ deviation from new physics, 
we find the allowed range
\begin{eqnarray}
0.07 \leq R_{B\tau \nu}  = \frac{\mathcal{BR}(B_u \to \tau \nu)^{MSSM}}{
\mathcal{BR}(B_u \to \tau \nu)^{SM}} \leq 2.51. \label{btaunubound:eq}
\end{eqnarray}

For the $R_{l23}$ constraint in Eq.~(\ref{rl23:eq}), 
Ref.~\cite{Antonelli:2008jg} finds that 
\bea
0.990 \leq R_{l23} \leq 1.018 \label{rl23bounda:eq}
\eea
when they use the value of $f_K/f_{\pi} = 1.189 \pm 0.007$ from 
Ref.~\cite{Follana:2007uv}. However if we use the average value of 
$f_{K}/f_{\pi} = 1.19 \pm 0.015$ from Ref.~\cite{lellouch}
\bea
0.96 \leq R_{l23} \leq 1.05. \label{rl23bound:eq}
\eea 
In this paper we will use this more conservative limit on the $R_{l23}$ rather
than the one in Eq.~(\ref{rl23bounda:eq}). From Eq.~(\ref{rl23_delta:eq}) it 
is clear that the more restrictive bound in Eq.~(\ref{rl23bounda:eq}) 
strongly disfavors the regions of low $M_A$ and large $\tan \beta$ that are 
allowed by the $B_u \to \tau \nu$ constraint in Eq.~(\ref{btaunubound:eq}), 
where the charged Higgs contribution to the $B_u \to \tau \nu$ process 
dominates that of the Standard Model.

The $B_s \to \mu^+ \mu^-$ rare decay has yet to be experimentally observed. 
The present experimental exclusion limit at 95\% C.L. from 
CDF~\cite{Bernhard:2005yn} is
\begin{eqnarray}
\mathcal{BR}(B_s\rightarrow \mu^+ \mu^-) \leq 5.8 \times 10^{-8},
\label{bsmumubound:eq}
\end{eqnarray}
which can put strong restrictions on possible flavor changing neutral currents 
in the MSSM at large $\tan\beta$. Additionally the projected exclusion
limit, at 95\% C.L., on this process for 4~fb$^{-1}$ at the Tevatron 
is~\cite{TeVBstomumu} 
\begin{eqnarray}
\mathcal{BR}(B_s\rightarrow \mu^+ \mu^-) \leq 2.8 \times 10^{-8}.
\label{bsmumuboundf:eq}
\end{eqnarray}
For the LHC, the projected ATLAS bound at 10~fb$^{-1}$ is~\cite{LHCBstomumu}
\begin{eqnarray}
\mathcal{BR}(B_s\rightarrow \mu^+ \mu^-) \leq 5.5 \times 10^{-9}. 
\label{bsmumuboundLHC:eq}
\end{eqnarray}
In addition, LHCb has the potential to claim a $3 \sigma \; (
5 \sigma)$ evidence (discovery) of a standard model signature with as little
as $\sim 2$fb$^{-1}$($6$fb$^{-1}$) of data~\cite{LHCB}.

\subsection{Direct dark matter detection constraints.}

As discussed in Section~\ref{dmth:sec} the spin-independent scattering 
cross-section for a neutralino off a heavy nucleus scales as $\tan^2 
\beta/M_A^4$ and therefore puts strong constraints on the SUSY parameter space.
At present the CDMS~\cite{Ahmed:2008eu} and  XENON~\cite{Angle:2007uj} 
collaborations have put a limit on the spin-independent neutralino-nucleon 
cross-section that is of the order of $10^{-7}$~pb~\cite{dmtools}. We will use 
the CDMS limits throughout this paper because we only consider 
neutralino masses greater than $100$~GeV and the current limits from the CDMS 
experiment are slightly stronger than those from XENON~\cite{dmtools} for 
these range of masses.  By the end of 2009, the 
sensitivity of the XENON100 experiment will improve by an order
of magnitude~\cite{xenon100}, while the sensitivity of the SuperCDMS experiment
will improve by factor of 5~\cite{Bruch:2007zz}.
Therefore in the near future, 
these direct dark matter detection experiments will be able to probe regions 
of SUSY parameter space that will also be probed by the Tevatron in 
non-standard Higgs searches.

\subsection{Parametric scenarios}

Due to the dependence on the messenger scale $M$
we consider different scenarios to illustrate the
interplay between B-physics, Higgs physics and dark matter searches within the
framework of the MSSM with large $\tan \beta$. In this section we will
assume the gaugino unification relation $|M_2| \simeq 2|M_1|$ and
that all the gauginos have equal phases. 
We shall consider 
four parameters: the CP-odd Higgs mass $M_A$, the ratio
of the two Higgs vacuum expectation values $\tan \beta$, the Higgsino mass 
parameter $\mu$ and $X_t = A_t - \mu/\tan \beta$, where $A_t$ is the 
stop-Higgs trilinear coupling. In order to consider the direct
dark matter detection constraints, we will also 
assume that the total relic density agrees with that found by WMAP, 
independently of the squark, neutralino and Higgs spectrum. This may require
an appropriate slepton spectrum or a departure from the standard thermal
dark matter predictions~\cite{Gelmini:2006pq}. In addition we considered
the non-standard Higgs boson search limits in 
Ref.~\cite{tevHA,conway,Carena:2005ek} and used the CPsuperH~\cite{Lee:2003nta}
program to project these constraints onto the $M_A\-- \tan \beta$ plane. 
Even though the Tevatron will be collecting about $8$~fb$^{-1}$ of data,
the projected $4$~fb$^{-1}$ CDF limit provides a conservative estimate
of the final reach of the Tevatron in the $H/A \to \tau \tau$ channel
because a realistic treatment of the detector and efficiencies may lead to 
somewhat weaker constraints than those shown in Ref.~\cite{conway}. The future 
LHC constraints on the $H/A \to \tau \tau$ channel, from 
Ref.~\cite{Carena:2005ek}, correspond to the projected limits at ATLAS for 
$30$~fb$^{-1}$ of data.

The good agreement between the Standard Model prediction and the experimental 
measurement of the branching ratio of $b \to s\gamma$ implies that either there
is some cancellation between the dominant new physics leading order Wilson 
coefficients in 
Eq.~(\ref{bsgh+:eq}) and Eq.~(\ref{bsgchi:eq}) or each of them 
are individually small. If $A_t$ is 
sizable and the sign of $\mu A_t$ is negative then a cancellation between the
chargino-stop and charged Higgs Wilson coefficients 
is possible for large enough
values of $\tan \beta$. For negligible values of $A_t$ the chargino-stop 
contribution
is suppressed, which requires the charged Higgs amplitude to be small. This 
suppression of the charged Higgs Wilson coefficient can be achieved by making 
the bottom-top-charged-Higgs vertex small through a cancellation between the
tree-level coupling and the one loop $\tan \beta$ enhanced correction in
Eq.~(\ref{bsgh+:eq}). In addition to the leading order contribution we have
also included the next-to-leading order contributions due to the charged Higgs 
as discussed in Ref.~\cite{Ciuchini:1997xe}. Additionally there is also a 
possible gluino contribution that can be significant if the squark masses are 
splitted. The gluino contribution, for 
the scenario in which $M \simeq M_{GUT}$, is relevant for large $\mu$, $M_3$ 
and $\tan \beta$ and depending on the sign of $\mu M_3$ this contribution 
interferes constructively or destructively with the other two contributions. 

Similarly to 
the $b\to s\gamma$ rare decay, the $B_s \to \mu^+ \mu^-$ rare decay is
sensitive to the scale $M$ at which supersymmetry breaking is communicated to 
the visible sector. 
Depending on the scale $M$ the structure of the $X_{RL}^{32}$ couplings
in Eq.~(\ref{xrl:eq}) and Eq.~(\ref{xrlnmfv:eq}) is different. If $M \sim 
M_{SUSY}$, this coupling is suppressed if 
$\epsilon_Y$ or equivalently $A_t$ is small compared to 
$m_{0}$ or $M_3$. However if $M$ is large compared to 
$M_{SUSY}$ then splittings in the squark masses, due to RG evolution, can 
induce a suppression of this coupling, due to a cancellation between the 
stop-chargino and sbottom gluino loops.

\begin{figure}
\begin{center}
\resizebox{9.cm}{!}{\includegraphics{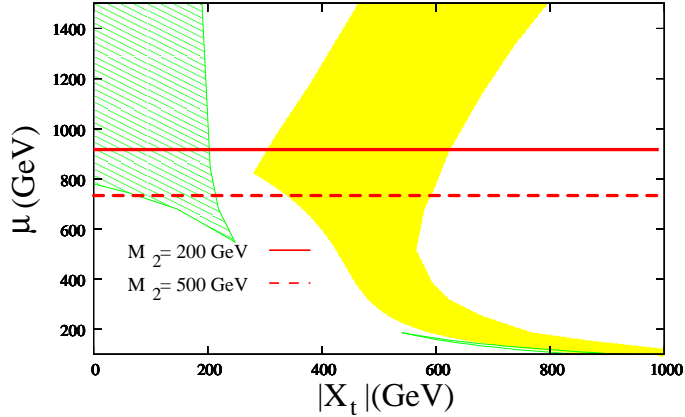}}
\end{center}
\caption{Plot of $X_t$ versus $\mu$, where $M_A=110$~GeV, $\tan \beta = 40$, 
sign($X_t$) negative, $M_3 =800$~GeV. The green (grey) hatched 
region is allowed by the $B_u \to \tau \nu$, $b \to s \gamma$ and $B_u \to 
\mu^+ \mu^-$ constraints for $M \sim M_{SUSY}$ while the yellow (light grey) 
region is allowed by the same constraints for the $M \simeq M_{GUT}$ scenario. 
The region below the horizontal red (dark grey) lines 
has been probed by the CDMS direct dark matter detection experiment assuming 
that the LSP is mainly bino and $|M_1|=2|M_2|$. The solid (dashed) line 
corresponds to a Wino mass parameter $M_2=200 \; (500)$~GeV.}
\label{ma110tb40:fig}
\end{figure}

The $B_u \to \tau \nu$ bound in Eq.~(\ref{btaunubound:eq}) imposes a 
complementary
constraint to that of $b \to s \gamma$. The lower bound of $R_{b\tau \nu}$
implies that there cannot be complete destructive interference between the 
SUSY and Standard Model contributions. Therefore there are two disconnected 
allowed regions in the $M_A \--\tan \beta$ plane: one where the charged Higgs
induced amplitude dominates 
the Standard Model contribution and the other where the opposite happens.

In these kind of scenarios, Eq.~(\ref{sigma_si:eq}) suggests that the 
spin-independent dark matter 
scattering cross-section is quite sensitive to the amount of the Higgsino 
component in the lightest neutralino. At low values of $\mu$ there is 
a large Higgsino component to the lightest neutralino and hence a larger 
scattering cross-section
through t-channel CP-even Higgs bosons. For large values of $\mu$, 
instead, the 
Higgsino component is much smaller, and so the coupling of the neutralino to 
the Higgs is suppressed leading to a smaller cross-section. In addition, the 
sensitivity of direct dark matter detection experiments, like CDMS, to the 
spin independent cross-section decrease with increasing LSP mass for 
$m_{\chi_1} \gsim 50$~GeV. Observe, however that the CDMS constraint
assumes the neutralino to be the dark matter candidate while in low SUSY 
breaking scenarios the LSP is naturally the gravitino, and therefore these
constraints should not apply.

Throughout this section we set 
the uniform left-handed soft squark mass parameter for the first two 
generations to $m_{0} = 1.5$~TeV, the third generation soft squark mass 
parameters are $m_{U_3}=m_{Q_3}=1.2$~TeV and the uniform right-handed down 
squark soft mass parameter is $m_R=1.5$~TeV.  
This form of the left handed down squark soft masses has been chosen so as to 
mirror a 20\% splitting in the soft masses due to their renormalization group
evolution in $M \simeq M_{GUT}$ scenario, as is naturally the case whenever the
gaugino masses are of the same order as the scalar masses at the scale
$M$. We have chosen a value of the third generation squark 
\begin{figure}
\begin{center}
\resizebox{9.cm}{!}{\includegraphics{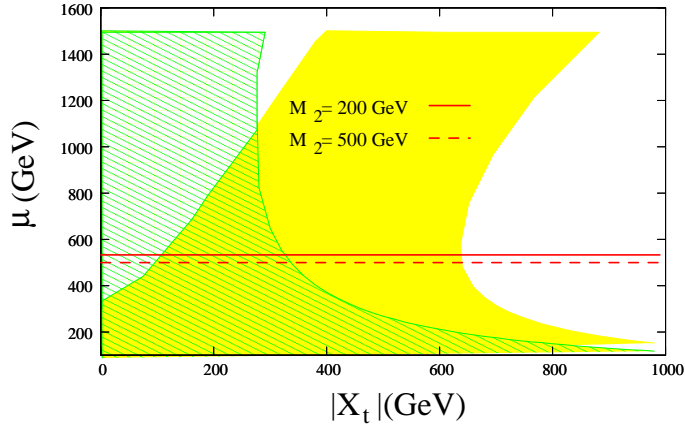}}
\end{center}
\caption{Plot of $X_t$ versus $\mu$, where $M_A=200$~GeV, $\tan \beta = 55$, 
sign($X_t$) negative, $M_3 =800$~GeV. The green (grey) hatched 
region is allowed by the $B_u \to \tau \nu$, $b \to s \gamma$ and $B_u \to 
\mu^+ \mu^-$ constraints for $M \sim M_{SUSY}$ while the yellow (light grey) 
region is allowed by the same constraints for the $M \simeq M_{GUT}$ scenario. 
The region below the horizontal red (dark grey) lines 
has been probed by the CDMS direct dark matter detection experiment assuming 
that the LSP is mainly bino and $|M_1|=2|M_2|$. The solid (dashed) line
corresponds to a Wino mass parameter $M_2=200 \; (500)$~GeV.}
\label{ma200tb55:fig}
\end{figure}
masses slightly
larger than 1~TeV, in order to satisfy the lightest CP-even Higgs mass
constraints for all the scenarios under study. 
For the $M \sim M_{SUSY}$ scenario we 
shall  assume a degenerate squark
spectrum with soft masses of 1.2~TeV. In this way, we can compare the
results with those in the case $M = M_{GUT}$ for which the
third generation masses, most relevant in the calculation of the
B-physics observables, have also values of 1.2~TeV.

In Fig.~\ref{ma110tb40:fig} and Fig.~\ref{ma200tb55:fig} 
we present the effects 
of the B physics constraints and the CDMS direct dark matter detection 
experiment limit on the $X_t \-- \mu$ plane for $(M_A,\tan \beta) = 
(110 \;
\mbox{GeV},40)$, and $(M_A,\tan \beta) = (200 \;
\mbox{GeV},55)$ 
respectively. These two sets of values
correspond 
to regions of parameter space which are close to being probed at the Tevatron 
in inclusive $A/H \to \tau \tau$ searches~\cite{tevHA}.
The 
green (grey) hatched region is the one allowed by the $b \to s\gamma$, $B_s \to
\mu^+ \mu^-$ and 
$B_u \to \tau \nu$ constraints for $M \sim M_{SUSY}$, while the yellow (light 
grey) region is allowed by the same constraints for $M \simeq 
M_{GUT}$ scenario.
The $B_u \to \tau \nu$ constraint does not depend on the 
parameter $X_t$ and therefore the constraint in Eq.~(\ref{btaunubound:eq})
selects a horizontal band in the $X_t \-- \mu$ plane. The regions below the
solid and dashed red (dark grey) lines is excluded by CDMS, 
for $M_2 = 200$~GeV and $M_2 = 500$~GeV, respectively.

In Fig.~\ref{ma110tb40:fig}, for $M_A = 110$~GeV and $\tan \beta=40$, 
The extra  gluino contributions to the $b \to s\gamma$ and, 
most relevantly, the $B_s \to \mu^+ \mu^-$ rare
decay rates, leads to a modification of the preferred values of $X_t$. 
While 
the B-physics constraints
lead to a  
preference for small values of $X_t$ in the $M \sim M_{SUSY}$ scenario, 
moderate values of $X_t \sim -500$~GeV are preferred in the $M \sim M_{GUT}$
case. 
Assuming that the LSP is the neutralino, which is natural in the
$M \sim M_{GUT}$ scenario,
the recent CDMS limits~\cite{Ahmed:2008eu} are quite strong and exclude 
regions below 
\begin{figure}
\begin{center}
\resizebox{9.cm}{!}{\includegraphics{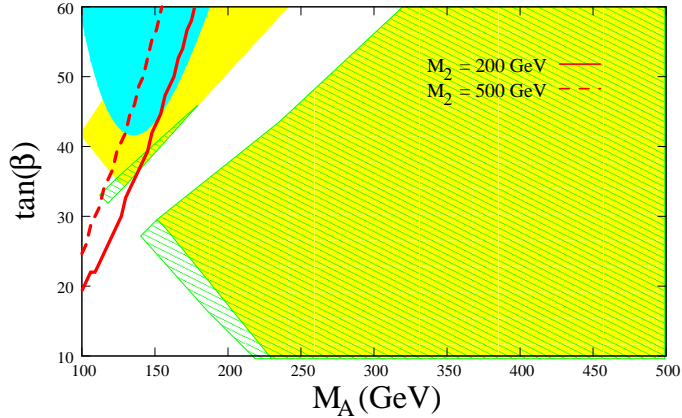}}
\end{center}
\caption{Plot of $M_A$ versus $\tan \beta$, where $X_t=-400$~GeV, 
$\mu=800$~GeV and $M_3 =800$~GeV. The green (grey) hatched 
region is allowed by the $B_u \to \tau \nu$, $b \to s \gamma$ and $B_u \to 
\mu^+ \mu^-$ constraints for $M \sim M_{SUSY}$ while the yellow (light grey) 
region is allowed by the same constraints for the $M \simeq M_{GUT}$ scenario. 
The region above the red (dark grey) lines 
has been probed by the CDMS direct dark matter detection experiment assuming 
that the LSP is mainly bino and $|M_1|=2|M_2|$. The blue-green region is 
excluded by Non-standard Higgs boson searches in the inclusive $\tau \tau$ 
channel at $1.8$fb$^{-1}$. The region above the black solid (dashed) lines
will be probed in the $H/A \to \tau \tau$ channel at the Tevatron (LHC) with
a luminosity of $4$~fb$^{-1}$ ($30$~fb$^{-1}$).}
\label{xt4mu8:fig}
\end{figure}
$|\mu| \sim 900$~GeV. Indeed, for these values of 
$M_A$ and $\tan \beta$ we observe that
large to moderate values of $\mu$ are preferred
for both SUSY breaking scenarios.

Fig.~\ref{ma200tb55:fig} shows the situation for $M_A=200$~GeV and $\tan \beta
= 55$. 
Similar to Fig.~\ref{ma110tb40:fig} small 
values of $X_t$ are preferred in the $M \sim M_{SUSY}$ scenario while moderate 
values of $X_t \sim -400$~GeV are preferred in the $M \sim M_{GUT}$ scenario.
For these values of $M_A$ and $\tan \beta$, the CDMS experimental 
bound is less stringent than in Fig.~\ref{ma110tb40:fig}, restricting
values below $|\mu| \sim 500$~GeV in the $M \simeq M_{GUT}$ scenario. 
Similar to Fig.~\ref{ma110tb40:fig},
the $B_s \to \mu^+ \mu^-$ constraint is the main discriminant between the
$M = M_{GUT}$ and $M = M_{SUSY}$ scenarios.



%
From Figs.~\ref{ma110tb40:fig} and Fig.~\ref{ma200tb55:fig}  we can observe 
some 
generic features. Independent of the SUSY breaking scale, for these regions of 
parameter space, that can be probed at 
the Tevatron, one obtains that $X_t \lsim 0.5 M_{SUSY}$. These low
values of the stop-mixing parameter $X_t$ imply an upper
bound on the lightest CP-even Higgs boson mass, $m_h \lsim 120$~GeV and
therefore could be within the reach of 
the Tevatron collider. 
As we had previously emphasized 
the regions close to $(M_A, \tan \beta) = (110 \mbox{ GeV}, 40)$ and $(200 
\mbox{ GeV}, 55)$ are yet to be probed at the Tevatron in $H/A \to \tau \tau$
searches at $1.8$~fb$^{-1}$. Using Figs.~\ref{ma110tb40:fig} 
and~\ref{ma200tb55:fig} we see that the region of parameter space around
$(M_A, \tan \beta, X_t, \mu) \sim (110 \mbox{ GeV}, 40, 0, 1 \mbox{ TeV})$ and 
$(M_A, \tan \beta, X_t, \mu) \sim (200 \mbox{ GeV}, 55, 0, 1 \mbox{ TeV})$ 
satisfies all the constraints in the $M \sim M_{SUSY}$ scenario. In addition, 
from Fig.~\ref{ma200tb55:fig}, we also find that the 
region of parameter space close to $(M_A, \tan \beta, X_t, \mu) \sim (200 
\mbox{ GeV}, 55, -400 \mbox{ GeV}, 800 \mbox{ GeV})$ satisfies 
all constraints for $M \sim M_{GUT}$. As the constraints from 
B-physics, Higgs physics and direct
dark matter searches get stronger this kind of analysis could help us
identify regions of parameter space that would still be 
compatible with all experimental limits.

In Fig.~\ref{xt4mu8:fig} we consider the $\mu=800$~GeV and $X_t=-400$~GeV 
parametric scenario. The region above the solid and dashed red (dark grey) 
lines has been  probed by CDMS in direct dark matter detection experiments,
for a Wino mass parameter $M_2 = 200$~GeV and $M_2 = 500$~GeV respectively. The
blue-green (medium grey) region is excluded by CDF and D0 in non-standard 
Higgs boson searches in the $\tau \tau$ channel at 1.8~fb$^{-1}$. The green 
(grey) hatched region is 
allowed by the experimental constraints on the $b \to s\gamma$, $B_s \to \mu^+ 
\mu^-$ and $B_u \to \tau \nu$ rare B decays for  
$M \sim M_{SUSY}$ while the yellow (light grey) region corresponds to the same
constraints for the $M \simeq M_{GUT}$ scenario. The region above the black 
solid (dashed) lines
\begin{figure}
\begin{center}
\resizebox{9.cm}{!}{\includegraphics{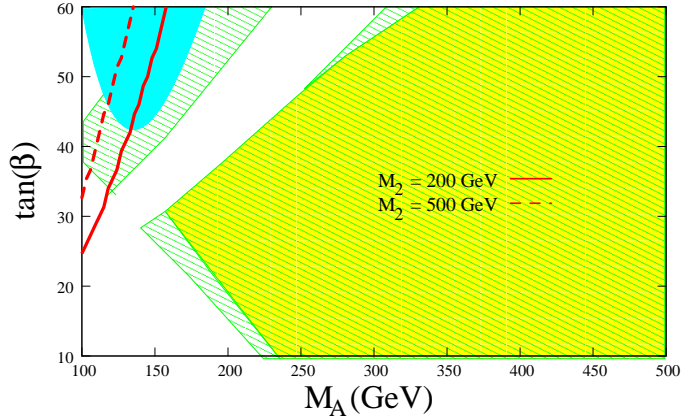}}
\end{center}
\caption{Plot of $M_A$ versus $\tan(\beta)$, for $X_t=0$, $\mu=1000$~GeV and 
$M_3 = 800$~GeV. The green (grey) hatched 
region is allowed by the $B_u \to \tau \nu$, $b \to s \gamma$ and $B_u \to 
\mu^+ \mu^-$ constraints for $M \sim M_{SUSY}$ while the yellow (light grey) 
region is allowed by the same constraints for the $M \simeq M_{GUT}$ scenario. 
The region above the red (dark grey) lines 
has been probed by the CDMS direct dark matter detection experiment assuming 
that the LSP is mainly bino and $|M_1|=2|M_2|$. The blue-green region is 
excluded by Non-standard Higgs boson searches in the inclusive $\tau \tau$ 
channel at $1.8$fb$^{-1}$.The region above the black solid (dashed) lines
will be probed in the $H/A \to \tau \tau$ channel at the Tevatron (LHC) with
a luminosity of $4$~fb$^{-1}$ ($30$~fb$^{-1}$).}
\label{xt0mu10:fig}
\end{figure}
will be probed in the $H/A \to \tau \tau$ channel at the Tevatron (LHC) with
a luminosity of $4$~fb$^{-1}$ ($30$~fb$^{-1}$).
In this parametric scenario, with $\mu X_t < 0$ and $\mu M_3 > 0$ and  
the different $b \to s\gamma$ contributions tend to cancel 
against each other. The chargino-stop approximately cancels the charged Higgs
contribution in the $M \sim M_{SUSY}$ scenario, while in the $M \simeq 
M_{GUT}$ scenario the chargino contribution tends to cancel both the charged
Higgs and the gluino contributions. There are two regions
allowed by $B_u \to \tau \nu$ constraint. The allowed region at low $M_A$
and large $\tan \beta$ is where the supersymmetric contribution dominates,
while in the large $M_A$ and low to moderate $\tan \beta$ region the Standard
Model contribution dominates. The $B_s \to \mu^+ \mu^-$ constraint
is quite strong for the $M \sim M_{SUSY}$ scenario due to a non-zero
$X_t$, excluding most of the region where the charged Higgs contribution
to $B_u \to \tau \nu$ becomes dominant,
while for the $M \simeq M_{GUT}$ a cancellation is induced in the 
flavor violating effects due to the splitting of the 
left-handed down squark soft mass parameters.

In Fig.~\ref{xt4mu8:fig} 
we see that somewhat smaller values of 
$\tan\beta$ are allowed in the $M \sim M_{SUSY}$ case than in the
$M \sim M_{GUT}$ case. This effect may be explained by the gluino
contributions to $b \to s \gamma$ : For positive values  of $\mu M_3$ 
a larger chargino-stop
contribution is required to get agreement with the experimental values, 
which may be obtained
for larger values of $\tan\beta$.   
The $R_{l23}$ constraint in Eq.~(\ref{rl23bound:eq}) becomes
too weak to give any significant constraint in Fig.~\ref{xt4mu8:fig} and
in the other scenarios we consider in this paper. Let us stress again, if 
we had considered the more restrictive bound on $R_{l23}$ in 
Eq.~(\ref{rl23bounda:eq}) the region of low $M_A$ and large $\tan \beta$ 
allowed by the other flavor constraints would be strongly disfavored.

As Fig.~\ref{ma110tb40:fig} suggested the region of small $M_A$
and $\tan \beta \sim $ 35--45 is allowed by all 
the B physics constraints and has yet to be probed by the Tevatron in inclusive
$A \to \tau \tau$ searches. The allowed region is larger for $M \sim M_{GUT}$
than for $M\sim M_{SUSY}$. For $M \sim M_{GUT}$, this region is however 
also constrained by direct  dark matter detection experiments like CDMS 
as can also be seen in  Fig.~\ref{xt4mu8:fig}. 
In addition, there is also a 
region around $(M_A,\tan \beta)=(200\mbox{ GeV},55)$ that  is allowed in the 
$M \simeq M_{GUT}$ scenario, which has yet to be probed in direct dark
matter detection experiments and non-standard Higgs searches.
However the
XENON100 and SuperCDMS experiments should be able to probe this region of 
parameter space in the near future due to their improved sensistivity.

In Fig.~\ref{xt0mu10:fig} we consider a scenario where $\mu=1$~TeV and 
$X_t=0$. 
The  chargino-stop contribution to the $b \to s \gamma$ rate 
is small in this scenario, because $A_t \sim 0$, while the charged 
Higgs contribution tends to be suppressed because of a cancellation 
between the 
1-loop and 2-loop contributions in Eq.~(\ref{bsgh+:eq}). The $B_s \to \mu^+ 
\mu^-$ constraint for the $M \simeq M_{GUT}$ scenario is strong because
there is no cancellation that occurs in Eq.~(\ref{xrlnmfv:eq}), while
it is weak in the $M \sim M_{SUSY}$ scenario because $A_t \sim 0$. In addition
there are two regions around $(M_A ,\tan \beta) = (175\mbox{ GeV}, 55)$ and 
$(M_A ,\tan \beta) = (115\mbox{ GeV}, 40)$ that are allowed by
all these constraints in the $M \sim M_{SUSY}$ scenario but disallowed
in the $M \simeq M_{GUT}$ scenario. As in the previous case, a significant
region of parameters consistent with all experimental appears at
low values of $M_A$ and large values of $\tan\beta$ but in this case 
is compatible with $M \sim M_{SUSY}$ and hence no CDMS restrictions
apply. In addition both Fig.~\ref{xt4mu8:fig} and Fig.~\ref{xt0mu10:fig}
also show that the Tevatron collider will be able to probe all these allowed 
regions in non-standard Higgs boson searches with a luminosity of $4$~fb$^{-1}
$. Futhermore, in both these scenarios, the LHC will be able to probe more
of the allowed regions of large $M_A$ and low $\tan \beta$, in the 
$H/A \to \tau \tau$ channel with $30$fb$^{-1}$ of luminosity.

\section{Conclusion} \label{concl:sec}

In this article we have studied the effect of varying the messenger scale
on B physics observables within Minimal Flavor Violating supersymmetric models.
In particular we found that the $b \to s\gamma$ and $B_s \to \mu^+ \mu^-$ rare
decays are sensitive to the scale $M$ at which supersymmetry 
breaking is communicated to the visible sector. Considering the
effects of the RG evolution which amounts to an alignment
of the left-handed squark masses with the up Yukawa couplings, with
uniform right-handed down squark soft masses and also uniform left-handed
down squark masses of the  first  two generations, we have
derived an analytic expression for the gluino contribution to the $b \to s
\gamma$ rare decay. We find that the gluino contribution is proportional
to the splitting between the third generation left-handed down squark mass 
and that of the first two generations. The relative sign of the gluino 
contribution to that of charged Higgs depends on the sign of $\mu M_3$. Hence 
in the case of the messenger scale $M \sim M_{GUT}$, when the splitting in the 
left-handed squark masses is non-zero, this contribution can be significant. 
In addition we also show the dependence of the dominant SUSY penguin 
contributions to the $B_s \to \mu^+ \mu^-$ rare decay on the scale $M$.

We have also studied the interplay between the B-physics constraints, 
dark matter direct detection experiments and non-standard Higgs boson searches 
at the Tevatron. For large soft squark masses, the spin-independent neutralino 
nucleon cross-section is proportional to $\tan^2 \beta/M_A^4$ and hence
direct detection experiments put strong constraints on regions of 
low $M_A$ ($M_A \simlt 200$~GeV)
and large $\tan \beta$. In particular, we have projected the CDMS direct dark 
matter detection experimental constraint on the $M_A \-- \tan \beta$ plane, 
for different values of $\mu$ and $M_2$. Independently of the messenger scale
$M$, the B physics, Higgs physics and Dark Matter 
experimental constraints suggest that
low values of $X_t$ and large to moderate values of $\mu$ are preferred.
Such low values of $X_t$ generally suggests an approximate upper bound on the 
lightest Higgs mass $m_h \lsim 120$~GeV, which may be within the reach of the 
Tevatron collider.

In addition,  we have presented parametric scenarios that  satisfy 
all  the B physics experimental constraints considered in this article,
within the  scenarios of low scale 
($M \sim M_{SUSY}$) and high scale ($M\simeq M_{GUT}$) supersymmetry breaking,
and can be probed by the Tevatron collider and the LHC in the near future. In 
particular for $M \simeq M_{GUT}$
we find a region  around $(M_A \simeq 200$~GeV and $\tan\beta \simeq 55$,
and moderate values of $|X_t|$ and $\mu$,
which is within the $4$fb$^{-1}$ reach  of the Tevatron collider
in non-standard Higgs searches. 
%
For $M \sim M_{SUSY}$, instead, smaller values of $|X_t|$ and moderate or
large values of $\mu$ are preferred, in order
to obtain acceptable values of $\mathcal{BR}(B_s \to \mu^+ \mu^-)$
and $\mathcal{BR}(b \to s \gamma)$. Moreover, we showed that for
$X_t \simeq 0$, there are large regions of parameter space for low
$M_A$ and large $\tan\beta$ that remain to be probed by non-standard Higgs
searches at the Tevatron collider. Apart from a region at similar values 
of $M_A$ and $\tan\beta$ as the
ones arising in the $M \sim M_{GUT}$ scenario,  we found an additional  
region, for smaller values of $M_A \simeq 115$~GeV and 
$\tan\beta ~\simeq 40$. This region only appears in scenarios with low
energy supersymmetry breaking, since it is constrained by direct dark 
matter  searches in the $M \simeq M_{GUT}$ scenario.

Our analysis suggests that, in Minimal Flavor Violating MSSM, future Higgs 
searches at the Tevatron and direct dark matter detection experiments at CDMS 
and XENON will reveal useful information about SUSY parameters and the scale of
supersymmetry breaking. For instance, the detection of a light non-standard 
Higgs boson at the Tevatron and dark matter at the XENON and CDMS experiments, 
by the end of 2009, would suggest a $M \sim M_{GUT}$ scenario, from which we 
can infer moderate values of $X_t$, large to moderate values of $\mu$ and a 
Standard Model Higgs boson mass $\lsim 120$~GeV. On the other hand, detection 
of a light non-standard Higgs boson and non-detection of dark matter may
suggest a lower SUSY breaking messenger scale, small values of $X_t$, large 
values of $\mu$ and a Standard Model Higgs boson mass close to that of the LEP 
experimental limit. 

\subsection*{\sc Acknowledgments}

Work at ANL is supported in
part by the US DOE, Div.\ of HEP, Contract DE-AC02-06CH11357. Fermilab
is operated by Fermi Research Alliance, LLC under Contract No.
DE-AC02-07CH11359 with the United States Department of Energy. A.M. was also
supported by MCTP and DOE under grant DE-FG02-95ER40899. We
would like to thank the Aspen Center for Physics and the KITPC, China,
where part of this work has been done.

\appendix
\section{Gluino contribution to $b \to s\gamma$} \label{append:1}

In the initial gauge basis the gluon-quark-squark interaction Lagrangian has 
the form
\bea
\mathcal{L}_{g} \supset \sqrt{2} g_3 \tilde{g}^a \left( (\tilde{d}_L^{*})^I T^a
d_L^I - (\tilde{d}_R^{*})^I T^a d_R^I \right).
\eea
Rotating the quarks into the mass basis by the matrices 
\bea
u_L^i \rightarrow U_L^{ij} u_L^j  \;\;\;\;
u_R^i \rightarrow U_R^{ij} u_R^j   \\
d_L^i \rightarrow D_L^{ij} d_L^j  \;\;\;\; 
d_R^i \rightarrow D_R^{ij} d_R^j 
\eea
and the down squarks by the matrices
\bea
\tilde{d}_L^I \rightarrow  U_L^{IJ} \tilde{d}_L^J \;\;\;\;
\tilde{d}_R^I \rightarrow  D_R^{IJ} \tilde{d}_R^J
\eea
so as to diagonalize the down squark soft masses. Hence the gluon-quark-squark 
interaction Lagrangian becomes
\bea
\mathcal{L}_g &\supset& \sqrt{2} g_3 \tilde{g}^a \left( (U_L^{\dagger} D_L)^{JI} 
(\tilde{d}_L^{*})^J T^a d_L^I - (\tilde{d}_R^{*})^I T^a d_R^I \right) \\
&=& \sqrt{2} g_3 \tilde{g}^a \left( (V_{CKM})^{JI} 
(\tilde{d}_L^{*})^J T^a d_L^I - (\tilde{d}_R^{*})^I T^a d_R^I \right).
\eea
However this rotation induces off-diagonal terms to the left-right and 
right-left blocks of the down-squarks, so that,
\bea
\mathcal{L}_{mass} &\supset& (\tilde{d}_L^*)^I (m_Q^2)^{I} (\tilde{d}_L)^I + 
(\tilde{d}_R^*)^I (m_R^2)^{I} (\tilde{d}_R)^I + \tilde{\mu}^* m_b 
(\tilde{d}_L^*)^I V_{CKM}^{I3} (\tilde{d}_R)^3 + h.c
\eea
where we have neglected terms proportional to the down and strange Yukawa's
and $\tilde{\mu} = \mu \tan \beta - A_b$. Hence all three of the left-handed
states mix with $m_{d_R}^3$. As $V_{CKM}^{33} \approx 1$ we explicitly 
diagonalize the $(\tilde{d}_L^3,\tilde{d}_R^3)$ sector so that
\bea
\tilde{d}_L^3 &\rightarrow& c_{\theta} \tilde{b}_1 - e^{-i\phi} s_{\theta} 
\tilde{b}_2\\
\tilde{d}_R^3 &\rightarrow& e^{i\phi} s_{\theta} \tilde{b}_1 + c_{\theta} 
\tilde{b}_2
\eea 
where
\bea
\phi &=& \arg(\tilde{\mu} ) \\
c_{\theta} &=& \cos \theta \\
s_{\theta} &=&  \sin \theta  \\
\cot(2\theta) &=& \frac{m_{Q_3}^2-m_R^2}{2|\tilde{\mu}| m_b }
\label{cot2t:eq} \\
m_{\tilde{b}_1}^2 &=& \cos^2 \theta m_{\tilde{Q}_3}^2 + \sin^2 \theta m_R^2 
+ 2 \cos \theta \sin \theta |\tilde{\mu} | m_b \\
m_{\tilde{b}_2}^2 &=& \sin^2 \theta m_{\tilde{Q}_3}^2 + \cos^2 \theta m_R^2 
- 2 \cos \theta \sin \theta |\tilde{\mu} | m_b
\eea
which leads to 
\bea
\mathcal{L}_{mass} &\supset& m_{0}^2\sum_{I=1}^2 (\tilde{d}_L^*)^I  
(\tilde{d}_L)^I + m_R^2 \sum_{I=1}^2 (\tilde{d}_R^*)^I (\tilde{d}_R)^I + 
+ m_{\tilde{b}_1}^2 \tilde{b}_1^* \tilde{b}_1 + m_{\tilde{b}_2}^2 
\tilde{b}_2^* \tilde{b}_2 \nonumber \\
& & \sum_{i=1}^2 \left(\tilde{\mu}^* m_b (\tilde{d}_L^*)^I V_{CKM}^{I3} 
(s_{\theta} \tilde{b}_1 + c_{\theta} \tilde{b}_2) +h.c \right) \\
\mathcal{L}_g &\supset& \sqrt{2} g_3 \tilde{g}^a \left( \sum_{i=1}^2 
\sum_{j=1}^3
V_{CKM}^{ij} \tilde{d}_L^{*i} T^a d_L^I - \sum_{i=1}^2 \tilde{d}_R^{*i} T^a 
d_R^i  \right. \nonumber \\
& & \left.+ \sum_{j=1}^3 V_{CKM}^{3j}  (c_{\theta} \tilde{b}_1^* - s_{\theta} 
e^{i \phi} \tilde{b}_2^*) T^a d_L^j - (s_{\theta} e^{-i\phi} \tilde{b}_1^* + 
c_{\theta} \tilde{b}_2^*) T^a d_R^3\right).
\eea
To leading order in the CKM matrix elements we
can further diagonalize the left-handed $I=1,2$ states so that
\bea
\tilde{d}_L^1 &=& \tilde{d}_1 + \frac{\tilde{\mu}^* m_b s_{\theta} e^{i\phi}}{
m_{b_1}^2-m_{0}^2} V_{CKM}^{13} \tilde{d}_3 +  \frac{\tilde{\mu}^* m_b 
c_{\theta}}{m_{b_2}^2-m_{0}^2} V_{CKM}^{13} \tilde{d}_6 \\
\tilde{d}_L^2 &=& \tilde{d}_2 + \frac{\tilde{\mu}^* m_b s_{\theta} e^{i\phi}}{
m_{b_1}^2-m_{0}^2} V_{CKM}^{23} \tilde{d}_3 +  \frac{\tilde{\mu}^* m_b 
c_{\theta}}{m_{b_2}^2-m_{0}^2} V_{CKM}^{23} \tilde{d}_6 \\
\tilde{b}_1 &=& \tilde{d}_3 + \frac{\tilde{\mu} m_b s_{\theta} e^{-i\phi}}{
m_{0}^2-m_{b_1}^2} V_{CKM}^{*13} \tilde{d}_1 +  \frac{\tilde{\mu} m_b s_{\theta} 
e^{-i\phi}}{m_{0}^2-m_{b_1}^2} V_{CKM}^{*23} \tilde{d}_2 \\
\tilde{d}_R^1 &=& \tilde{d}_4 \\
\tilde{d}_R^3 &=& \tilde{d}_5 \\
\tilde{b}_2 &=& \tilde{d}_6 + \frac{\tilde{\mu} m_b c_{\theta}}{
m_{0}^2-m_{b_2}^2} V_{CKM}^{*13} \tilde{d}_1 +  \frac{\tilde{\mu} m_b c_{\theta}
}{m_{0}^2-m_{b_1}^2} V_{CKM}^{*23} \tilde{d}_2
\eea
Hence
\bea
\mathcal{L}_g &\supset& \sqrt{2} g_3 \tilde{g}^a \left[ \sum_{i,j=1}^2
V_{CKM}^{ij} \tilde{d}_i^* T^a d_L^j + \frac{(m_{0}^2-m_R^2)(m_{0}^2-
m_{Q_3}^2)}{(m_{0}^2 - m_{b_1}^2)(m_{0}^2-
m_{b_2}^2)} \sum_{i=1}^2 V_{CKM}^{i3} \tilde{d}_i^* T^a d_L^3 \right.
\nonumber \\
& & \left(c_{\theta} - \frac{|\tilde{\mu}|m_b}{m_{b_1}^2-
m_{0}^2} s_{\theta}\right) \sum_{i=1}^2 V_{CKM}^{3j} \tilde{d}_3^* T^a d_L^j 
- e^{i\phi} \left(s_{\theta} + \frac{|\tilde{\mu}|m_b}{m_{b_1}^2-
m_{0}^2} c_{\theta}\right) \sum_{i=1}^2 V_{CKM}^{3j} \tilde{d}_6^* T^a d_L^j 
\nonumber \\ 
& & c_{\theta}  \tilde{d}_3^* T^a d_L^3 - e^{i\phi} 
s_{\theta}  \tilde{d}_6^* T^a d_L^3 - \frac{\mu^*m_b(m_{0}^2-
m_{Q_3}^2)}{(m_{0}^2-m_{b_1}^2)(m_{0}^2-m_{b_2}^2)} \sum_{i=1}^2 V_{CKM}^{i3}
\tilde{d}_i^* T^a d_R^3 \nonumber \\
& & \left. -(s_{\theta} e^{-i\phi} \tilde{d}_3^* T^a d_R^3 + c_{\theta}
\tilde{d}_6^* T^a d_R^3) -\tilde{d}_4^* T^a d_R^1 - \tilde{d}_5^* T^a d_R^2 
\right]
\eea
where we have neglected higher orders in the CKM matrix.
Therefore the vertex factors are 
\bea
\Gamma_{DL}^{ki} &=& \left\{\begin{array}{ll}
V_{CKM}^{ki} &  \mbox{$i=1,2$ ; $k=1,2$}\\
\frac{(m_{0}^2-m_R^2)(m_{0}^2-m_{Q_3}^2)}{(m_{0}^2 - m_{b_1}^2)(m_{0}^2-m_{b_2}^2)}
V_{CKM}^{k3} & \mbox{$i=3$ ; $k=1,2$} \\
\left(c_{\theta} - \frac{|\tilde{\mu}|m_b}{m_{b_1}^2-
m_{0}^2} s_{\theta}\right) V_{CKM}^{3i}  & \mbox{$i=1,2$ ; $k=3$} \\
c_{\theta}   & \mbox{$i=1,2$ ; $k=3$} \\
-e^{i\phi} (s_{\theta} + \frac{|\tilde{\mu}|m_b}{m_{b_1}^2-
m_{0}^2} c_{\theta}) V_{CKM}^{3i} & \mbox{$i=1,2$ ; $k=6$} \\
- e^{i\phi} s_{\theta}  & \mbox{$i=3$ and $k=6$} \\
0 & \mbox{otherwise}
\end{array} \right. \\
\Gamma_{DR}^{ki} &=& \left\{\begin{array}{ll}
\frac{\mu^*m_b(m_{0}^2-m_{Q_3}^2)}{(m_{0}^2-m_{b_1}^2)(m_{0}^2-m_{b_2}^2)} 
V_{CKM}^{k3} & \mbox{$i=3$ ; $k=1,2$} \\
s_{\theta} e^{-i\phi} & \mbox{$i=3$ and $k=3$} \\
\delta^{(k-3)i} & \mbox{$i=1,2$ and $k=4,5$} \\
c_{\theta} & \mbox{$i=3$ and $k=6$} \\
0 & \mbox{otherwise}
\end{array} \right.
\eea
Now rewriting the following identity in Eq.~(\ref{cot2t:eq})
\bea
c_{\theta} s_{\theta} (m_R^2 - m_{Q_3}^2) + |\tilde{\mu} |
m_b (c_{\theta}^2 - s_{\theta}^2) = 0
\eea
we have
\bea
c_{\theta}^2 (m_{b_1}^2 - m_{0}^2) - |\tilde{\mu} | m_b c_{\theta} 
s_{\theta} &=& c_{\theta}^2 (m_{Q_3}^2-m_{0}^2) + c_{\theta}^2 s_{\theta}^2 
(m_R^2 - m_{Q_3}^2) + \nonumber \\
& & c_{\theta} s_{\theta} |\tilde{\mu} | m_b (2
c_{\theta}^2-1) \\
&=& c_{\theta}^2 (m_{Q_3}^2-m_{0}^2) \\
s_{\theta} c_{\theta} (m_{b_1}^2 - m_{0}^2) - |\tilde{\mu} | m_b 
s_{\theta}^2 &=&  s_{\theta} c_{\theta} (m_{Q_3}^2-m_{0}^2) \\
s_{\theta}^2 (m_{b_2}^2 - m_{0}^2) + |\tilde{\mu} | m_b c_{\theta} 
s_{\theta} &=& s_{\theta}^2 (m_{Q_3}^2-m_{0}^2) \\
s_{\theta} c_{\theta} (m_{b_2}^2 - m_{0}^2) + |\tilde{\mu} | m_b 
c_{\theta}^2 &=&  s_{\theta} c_{\theta} (m_{Q_3}^2-m_{0}^2)
\eea
Hence using the amplitudes defined in Ref.~\cite{Bertolini:1990if} we find the 
Wilson coefficients
\bea
C_{7,8}^{\tilde{g}} &=& \frac{\sqrt{2} \pi \alpha_s}{G_f} 
\frac{M_3 e^{-i\phi}}{m_b}(m_{0}^2-m_{Q_3}^2) \left(
\frac{f_{\gamma,g}^5(x_{g0})}{m_{0}^2} 
\frac{|\tilde{\mu}|m_b}{(m_{0}^2-m_{b_1}^2)(m_{0}^2-m_{b_2}^2)} \right.\\
& & \left. + s_{\theta} c_{\theta} \left\{
\frac{f_{\gamma,g}^5(x_{g1})}{m_{b_1}^2 (m_{b_1}^2-m_{0}^2)}
- \frac{f_{\gamma,g}^5(x_{g2})}{m_{b_2}^2(m_{b_2}^2-m_{0}^2)}\right\} \right) \nonumber
\eea
where
\bea
f_{\gamma}^5(x) &=&\frac{-2-2x}{9(x-1)^2} + \frac{4x}{9(x-1)^3} \log x 
\label{fs:eq}\\
f_{g}^5(x) &=& \frac{13-5x}{3(x-1)^2} + \frac{x-9}{3(x-1)^3} \log x \nonumber
\eea

\end{document}